 \documentclass[final,5p,times,twocolumn]{elsarticle}

\usepackage{amsmath}

\usepackage[colorlinks=true, linkcolor=blue, citecolor=blue, urlcolor=blue]{hyperref}

\usepackage{amssymb}
\usepackage{lipsum}
\journal{Physics Letters B}

\begin{document}

\begin{frontmatter}



	\title{Gamow shell model description of neutron-rich He hyper-isotopes}
	

  \author[first]{Xin Li}
  
\author[second,third]{Nicolas Michel}

\ead{nicolas.michel@impcas.ac.cn}
\author[second,third]{Jianguo Li}
\author[first]{Xian-Rong Zhou}
\ead{xrzhou@phy.ecnu.edu.cn}

\affiliation[first]{
    organization={Department of Physics, East China Normal University},
    city={Shanghai},
    postcode={200241},
    country={China}
}

\affiliation[second]{
    organization={Heavy Ion Science and Technology Key Lab, Institute of Modern Physics, Chinese Academy of Sciences},
    city={Lanzhou},
    postcode={730000},
    country={China}
}

\affiliation[third]{
    organization={School of Nuclear Science and Technology, University of Chinese Academy of Sciences},
    city={Beijing},
    postcode={100049},
    country={China}
}

\begin{abstract}
	The Gamow shell model (GSM) framework is extended to the study of weakly bound hypernuclei. As an initial application, we investigate the neutron-rich He hyper-isotopes from {$^6_{\Lambda}$He} to {$^9_{\Lambda}$He}, accurately treating the unbound or loosely bound character of hypernuclear many-body states. The energy spectra, calculated using a phenomenological Hamiltonian with a effective $YN$ interaction, show good agreement with available experimental data. 
    In particular, neutron-emitting resonant states are predicted for the neutron-rich nuclei $^{5-8}$He and the hypernucleus $^6_{\Lambda}$He. Furthermore, neutron densities exhibit the long-range character of weakly bound and resonant states. This study demonstrates that GSM is a practical tool for describing the complex structure of  hypernuclei, especially for those close to drip lines.
    
\end{abstract}



\begin{keyword}
%% keywords here, in the form: keyword \sep keyword, up to a maximum of 6 keywords
Gamow shell model \sep Neutron-rich hypernuclei  \sep Resonance \sep $\Lambda N - \Sigma N$ coupling



\end{keyword}

\end{frontmatter}

%\tableofcontents


%%%%%%%%%%%%%%%%%%%%%%%%%%%%%%%%%%%%%%%%%%%%%%%%%%%%%%%%%%
%                    begin  introduction
%%%%%%%%%%%%%%%%%%%%%%%%%%%%%%%%%%%%%%%%%%%%%%%%%%%%%%%%%%

\section{Introduction}
	\label{intro}
	The hypernuclei close to the extreme regions of the nuclear chart are complicated many-body quantum systems with similar properties to exotic nuclei. Specifically, the addition of a $\Lambda$ hyperon to a well unbound nucleus typically results in a hypernucleus that is either weakly bound or narrowly resonant. This phenomenon originates from the absence of the Pauli exclusion principle between nucleons and $\Lambda$ hyperons, which has been noticed by Dalitz and Levi Setti in the early 60s \cite{Dalitz1963}. Indeed, due to this specific feature, the $\Lambda$ hyperon mainly occupies the bound $0s_{1/2}$ orbital. Therefore, the addition of a single hyperon significantly increases the binding energy to the newly formed hypernucleus \cite{PhysRevC.76.034312,GAL2013445}, which makes it possible to gain insights into the structure of broad hypernuclear many-body resonant states.
	
	Emulsion experiments have demonstrated the existence of light nuclei with large neutron excess, as reported in Ref.\cite{davis200550}. The synthesis of loosely bound or resonant hypernuclei is evidently linked to the development of facilities dedicated to the study of nuclei at drip lines (see Ref. \cite{feliciello2015experimental} for a review about the subject). Up to now, experimental facilities can reach the neutron drip line only up to  $A \sim$ 40 \cite{PhysRevLett.122.052501}. However, current
	and future experimental facilities, such as FAIR \cite{SAITO2012218}, RHIC \cite{PhysRevC.98.014910}, J-PARC \cite{NAGAE2008486c}, HIAF \cite{MA2017169} 
	% HIAF \cite{liu2017production} 
	or RIBF \cite{okuno2012progress}, aim to extend this limit higher on the nuclear chart \cite{BOTVINA20157}.
	
	Hyperons are unstable as they decay into nucleons and mesons \cite{RevModPhys.33.231,RevModPhys.88.035004}. However, the associated reactions originate from beta decay only, so that the lifetime of hyperons is about 10 orders of magnitude longer than that of nucleons in nuclear resonances \cite{gaillard1984hyperon}. Therefore, similarly to beta-unstable nuclei, hypernuclei can be considered as bound systems in terms of the nuclear interaction. In the domain of light hypernuclei, it is possible to experimentally observe helium hyper-isotopes with $A$ = 6 to 8 baryons \cite{davis200550}. Of particular interest is $_{\Lambda}^6$He, whose ground state exhibits an extended valence neutron density \cite{averyanov2008studying}. This arises because of the well unbound character of $^5$He, whose neutron-emission width is about 600 keV \cite{BROWN1979157}. In contrast, the addition of a $\Lambda$ hyperon to $^6$He leads to the disappearance of the neutron halo in $^7_{\Lambda}$He \cite{filikhin2009spectroscopy,PhysRevC.94.021302}. 
	However, heavier drip line hypernuclei are more difficult to synthesize. For that matter, Saha $et$ $al.$ performed a pioneering experiment, consisting a double charge exchange reaction ($\pi^-$, K$^+$) on a $^{10}$B target to produce $^{10}_\Lambda$Li. Significant discrete peaks were difficult to observe, nevertheless due to the small cross section obtained \cite{PhysRevLett.94.052502}. Similarly, the Japan Proton Accelerator Research Complex (J-PARC) collaboration plans to produce the $_{\Lambda}^9$He hypernucleus using the double-charge exchange reaction $(\pi^-,\mathrm{K}^+)$ on a $^{9}$Be target \cite{10.1093/ptep/pts056}. 
	
	The physics of $\Lambda$-hypernuclei bears a fundamental difference with nucleonic systems: the number of proton{s}, neutron{s}, $\Lambda$ and $\Sigma^{-,0,+}$ particles {are} not conserved, since only total charge, strangeness and baryon number are good quantum numbers. In practice, this arises due to the relative {strength} of the $\Lambda n$$-$$\Sigma^- p$ and $\Lambda p$$-$$\Sigma^+ n$ couplings \cite{gibson1994importance}.  Indeed, due to isospin conservation, the direct $\Lambda N$$-\Lambda N$ interaction contains no one-pion-exchange contribution, contrary to $\Lambda N-$$\Sigma N$ conversion channels.
	As a consequence, the $\Lambda N-$$\Lambda N$ interaction is comparatively weaker, so that fixed-baryon number fluctuations increase. Alternatively, the three-body $\Lambda NN-$$\Lambda NN$ interaction might be relatively important \cite{PhysRevLett.84.3539,gal2016strangeness}. The $\Lambda N-$$\Sigma N$  {conversion} couplings are also suspected to play an important role in the structure of heavier neutron-rich $\Lambda$ hypernuclei, whose total isospin increases along with $A$ \cite{PhysRevC.89.061302}. In fact, $\Lambda N$$-\Sigma N$ conversions are necessary to understand the binding energies of known hypernuclei \cite{PhysRevLett.84.3539}.
	
Whereas the theoretical apparatus devised for the study of drip line nuclei is now well developed \cite{PhysRevLett.95.042503, MA2020135673}, models aiming at describing weakly bound and unbound hypernuclei are scarce. E.~Hiyama et al. investigated the hypernuclear systems $_\Lambda^5$He to $_\Lambda^7$He, employing an $\alpha$ + $\Lambda$ + $n$ configuration and utilizing a phenomenological Hamiltonian defined in Jacobi coordinates \cite{PhysRevC.53.2075, PhysRevC.80.054321, hiyama2015resonant}. In Ref. \cite{PhysRevC.107.054302}, the energy spectra of $_{\Lambda}^6$He$-_{\Lambda}^9$He { were} studied by T.~Myo and E.~Hiyama with a similar approach within the {framework} of the cluster orbital shell model (COSM). These authors predicted one deeply bound state and two excited resonant states in the $_{\Lambda}^9$He hypernucleus. However, they did not consider $\Lambda N$$-\Sigma N$ conversions in their hyperon-nucleon ($YN$) interactions, but instead applied an effective single-channel $\Lambda N$ interaction.  The no-core shell model (NCSM) has also been applied to $_{\Lambda}^{5-11}$He hypernuclei~\cite{wirth2018}. In this approach, only low-lying states with natural parity were presented, and continuum effects were not included. As noted by the authors, these missing degrees of freedom can significantly impact states close to the neutron emission threshold, potentially lowering their absolute energies.

    In this work, we perform the first application of the Gamow Shell Model (GSM) to hypernuclei, extending the complex-energy configuration interaction framework to include strangeness degrees of freedom and explicit $\Lambda N - \Sigma N$ conversion channels. The GSM has proven effective in describing the exotic properties of drip line nuclei ~\cite{michel2008shell,hu2020ab,li2021recent,xie2023investigation}. Continuum coupling is directly included at basis level by using the Berggren basis, in which, bound, resonance, and continuum single-particle states are treated on an equal footing in the complex momentum plane. This capability makes it particularly suitable for hypernuclear systems near the drip lines, where continuum coupling plays a crucial role. In the present study, we construct an effective $YN$ interaction with minimal parameters and investigate the structure of neutron-rich He hyper-isotopes. The present study especially focuses on validating the applicability of the GSM framework to hypernuclear systems, thereby laying the foundation for future studies incorporating more realistic hyperon-nucleon interactions.
	
	This paper is organized as follows. The theoretical framework is
	outlined in Sec.~\ref{sec:the}, which contains a short overview of GSM, description of the GSM Hamiltonian within the COSM {framework}, and the optimization protocol of the effective Hamiltonian considered. The
results are presented in Sec.~\ref{sec:res}. Section~\ref{sec:sum} contains the summary and perspectives for future studies.

%%%%%%%%%%%%%%%%%%%%%%%%%%%%%%%%%%%%%%%%%%%%%%%%%%%%%%%%%%
%                    begin  Theoretical framework
%%%%%%%%%%%%%%%%%%%%%%%%%%%%%%%%%%%%%%%%%%%%%%%%%%%%%%%%%%
    
\section{Theoretical framework}
	\label{sec:the}
	\subsection{GSM general method}
	We will briefly outline the GSM formalism (for more details, see Ref.~\cite{michel2021gamow}). In this work, we assume that the neutron-rich He hyper-isotopes can be described by a system of $A_{\text{val}}$ valence baryons evolving around an inert $^4$He core, similarly to the calculation of He and Li isotopic and isotonic chains of Refs. \cite{PhysRevC.96.054316,PhysRevC.102.024309}. This allows us to express GSM in the {framework} of the cluster orbital shell model (COSM) \cite{PhysRevC.38.410}, which is translationally invariant and thus contains no center-of-mass (c.m.) motion. 
	
	The GSM Hamiltonian reads with COSM coordinates~\cite{PhysRevC.96.054316}
	\begin{equation}
		\centering
		\label{eq:1}
		\hat{H}_{\mathrm{GSM}}=\sum_{i=1}^{A_{\text {val }}}\left(\frac{{p}_i^2}{2 \mu_i}+\hat{U}_i^{(c)}\right)+\sum_{i<j}^{A_{\text {val }}}\left(\hat{V}_{i j}^{(r e s)}+\frac{{p}_i \cdot {p}_j}{M_c}\right),
	\end{equation}
	where $A_{\text{val}}$ is the number of valence baryons (nucleons and hyperons), and $\mu_i$, $M_{c}$ stand for the effective mass of the baryon and the mass of core, respectively. $\hat{U}_i^{(c)}$ is the single-particle core potential acting on the $i$-th particle, and $\hat{V}_{i j}^{(r e s)}$ is the interaction between valence nucleon-nucleon or nucleon-hyperon pairs. The kinetic operator at the end of Eq.~(\ref{eq:1}) is the recoil term generated from the motion of the core with respect to valence baryons. As the GSM Hamiltonian is directly defined with COSM coordinates, it is straightforward to optimize its parameters so that they reproduce experimental data \cite{li2021recent,xie2023investigation}.

	The GSM Hamiltonian takes the form of a matrix to diagonalize when represented with Slater determinants. The single-particle (s.p.) basis utilized to devise the basis of Slater determinants is the Berggren basis \cite{berggren1968use}, generated by a finite-depth potential (typically of Woods-Saxon type). The Berggren basis contains bound, resonance, and scattering states, and satisfies the following completeness relation for a given partial wave of $(\ell,j)$ quantum numbers:
	\begin{equation}
		\centering
		\label{eq:2}
		\begin{aligned}
			& \sum_{n=b, d}\left|\widetilde{u}_n\right\rangle\left\langle u_n\right| 
			& +\int_{\mathcal{L}^{+}}|\widetilde{u}(k)\rangle\langle u(k)|  dk = 1,
		\end{aligned}
	\end{equation}
	where $b$ and $d$ are bound states and selected decaying resonant states, respectively.  $\mathcal{L}^{+}$ is the complex contour of scattering states, which must include the resonance states present in the discrete sum. Any s.p.~bound state can be described in this basis, as well as the decaying resonances of complex momentum $k$ located between the contour $\mathcal{L}^{+}$ and the real axis \cite{
		berggren1968use, PhysRevC.47.768}. In practice, the contour is truncated at a $k_{\text{max}}$  value, typically $2-3$ fm$^{-1}$, and discretized, so that only a finite number of s.p.~states are present in numerical calculations. The basis of Slater determinants is therefore immediate to generate through the occupation of all possible one-body states from every nucleon or hyperon partial wave. Due to the inclusion of resonances and scattering states of complex momentum, the GSM Hamiltonian matrix is complex symmetric, and, hence, possesses complex eigenvalues. Eigenenergy and width are then provided by the real and imaginary parts of the Hamiltonian eigenvalues, respectively.
	Complex-symmetric matrices are diagonalized by way of the complex extension of the Jacobi-Davidson method \cite{Sleijpen1996jacobi}. Resonance states are identified in the complex energy spectrum of unbound states formed by both resonance and scattering states with the overlap method \cite{PhysRevC.102.024309,PhysRevC.104.024319}. 
	
	\subsection{GSM effective Hamiltonian of hypernuclei}
	
	The GSM interaction {consists} of a one-body core-valence potential and a two-body interaction acting between valence baryons. The core-valence potential is taken as a Woods-Saxon field, with a central part, a spin-orbit part, and a Coulomb part acting on protons and charged hyperons \cite{michel2008shell,PhysRevC.96.054316}:
	\begin{equation}
		\centering
		\label{eq:3}
		\begin{aligned}
			U_{\mathrm{c}}(r)=V_0 f(r)-4 V_{\ell s} \frac{1}{r} \frac{d f(r)}{d r} (\boldsymbol{\ell} \cdot \boldsymbol{s}) + U_{\mathrm{Coul}}(r),
		\end{aligned}
	\end{equation}
	where $f(r)=-\left( 1+\exp \left[ (r-R_0)/a \right] \right)^{-1}$ is the standard WS form factor, $V_0$ is the central potential depth, $R_0$ is the WS radius, $V_{\ell s}$ is the spin-orbit strength, and $a$ is the WS potential diffuseness. The latter parameters are {applied to} all valence protons, neutrons and hyperons. The Coulomb potential is generated by a spherical Gaussian charge distribution for the $^4$He core, whose charge radius is taken at its experimental value of $R_\mathrm{ch}$ = 1.681 fm. The nucleon-nucleon ($NN$) interaction is modeled as a sum of central ($V_c$), spin–orbit ($V_{LS}$), tensor ($V_T$), and Coulomb ($V_{\text{Coul}}$) components:
\begin{equation}
	\label{eq:4}
	V = V_c + V_{LS} + V_T + V_{\text{Coul}}.
\end{equation}

The central, spin–orbit, and tensor terms follow the Furutani–Horiuchi–Tamagaki (FHT) form \cite{michel2021gamow,10.1143/PTP.62.981}:
\begin{equation}
	\label{eq:5}
	V_c(r) = \sum_{n=1}^3 V_c^n \left(W_c^n + B_c^n \hat{P}_\sigma - H_c^n \hat{P}_\tau - M_c^n \hat{P}_\sigma \hat{P}_\tau\right) e^{-\beta_c^n r^2},
\end{equation}
\begin{equation}
	\label{eq:6}
	V_{LS}(r) = \boldsymbol{L} \cdot \boldsymbol{S} \sum_{n=1}^2 V_{LS}^n \left(W_{LS}^n - H_{LS}^n \hat{P}_\tau\right) e^{-\beta_{LS}^n r^2},
\end{equation}
\begin{equation}
	\label{eq:7}
	V_T(r) = S_{ij} \sum_{n=1}^3 V_T^n \left(W_T^n - H_T^n \hat{P}_\tau\right) r^2 e^{-\beta_T^n r^2},
\end{equation}
where $r$ is the relative distance between two baryons, $\boldsymbol{L}$ is their relative orbital angular momentum, $\boldsymbol{S}=\left(\sigma_i+\sigma_j\right) / 2, S_{i j}=3\left(\sigma_i \cdot \hat{r}\right)\left(\boldsymbol{\sigma}_j \cdot \hat{r}\right)-\sigma_i \cdot \sigma_j$, and $\hat{P}_\sigma \text { and } \hat{P}_\tau$ are spin and isospin exchange operator, respectively.

The nucleon-nucleon and the core–neutron interactions used in this work are the same as those in Ref.~\cite{PhysRevC.96.054316}, which successfully describe both bound and unbound nuclear systems with $A \leqslant 9$. For the core-$\Lambda$ interaction, we adjust the potential depth $V_0$ to reproduce the experimental $\Lambda$ binding energy $B_\Lambda = 3.12$ MeV observed in $^5_\Lambda$He, with an s-wave configuration for the $\Lambda$ hyperon. The hyperon–nucleon interaction is constructed following the symmetry-based approach as in Refs.~\cite{POLINDER2006244,haidenbauer2013baryon,petschauer2020hyperon}, which connects the $NN$ and $YN$ sectors through a limited set of effective coupling constants. In the present work, only the dominant coupling constants ($C^{8a}$ and $C^{8s}$) ~\cite{POLINDER2006244} are adjusted to reproduce the $1/2^+$ and $3/2^+$ states in $^7_\Lambda$He. Remaining parameters are kept fixed due to their negligible impact on the calculated observables. Although these $NN$ and $YN$ interactions are effective, they provide a practical and flexible tool within the GSM framework to study light hypernuclei. The present framework focuses on reproducing the low-energy spectra of hypernuclei with simplified effective $YN$ interactions and minimal tuning of phenomenological inputs.

	\begin{figure*}[htbp]
		\begin{center}
			\includegraphics[width=17.0 cm]{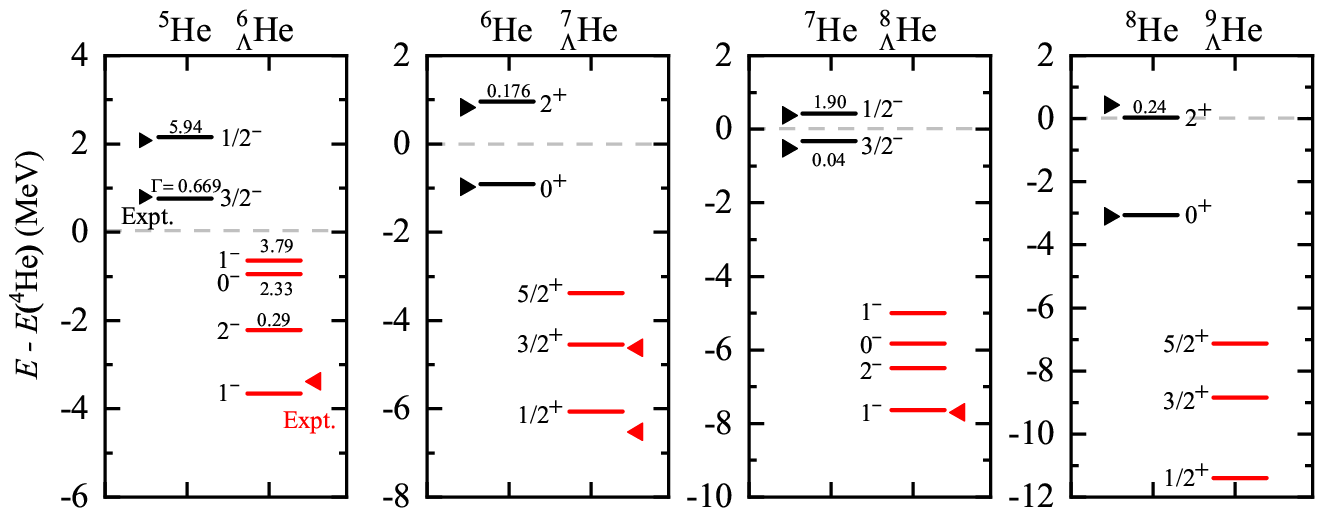}~
			\caption{The calculated energy spectra of the He isotope nuclei
				and hypernuclei. Lines are calculated values
				and triangles are experimental data. The experimental data are taken
				in Refs. \cite{tilley2002energy,tilley2004energy,PhysRevC.73.044301,korsheninnikov1993experimental}. Black and red represent the nuclei and hypernuclei, respectively. Widths are provided in MeV above the energy of the resonance eigenstate. }\label{figure1}
		\end{center}
	\end{figure*}

	\begin{table*}[]
		\caption{The energies $E$ (in MeV) and hyperon binding energies $B$ (in MeV) of the low-lying states of $^6_{\Lambda}$He $- $ $^9_{\Lambda}$He together with those of the corresponding states of $^5$He $-$ $^8$He, respectively. The energies in parentheses are decay
			widths (in MeV) for resonance states. For comparison, we also show the results of GSM without considering $\Lambda N-\Sigma N$ channel coupling ($E_\mathrm{GSM^*}$) and the results of the cluster orbital shell model ($E_\mathrm{Myo}$) \cite{PhysRevC.107.054302}.}\label{table:one}
		\renewcommand{\arraystretch}{1.3}
		\setlength{\tabcolsep}{1.0mm}{
			\begin{tabular}{cccccccccccccccccc}
				\hline\hline
				Hypernucleus & state &  & $E_\mathrm{exp}$ &  & $E_\mathrm{GSM}$ &  & $E_\mathrm{GSM^{\ast}}$ &  & $E_\mathrm{Myo}$ &  & $B_\mathrm{exp}$ &  &$B_\mathrm{GSM}$ &  & $B_\mathrm{GSM^{\ast}}$ &  & $B_\mathrm{Myo}$ \\ \hline
				$^5$He          & 3/2$^-$  &  &    0.798(0.648)     &  &    0.759(0.669)   &  &     &  &    0.75(0.59)  &  &    &  &     &  &       &  &      \\
				& 1/2$^-$  &  &   2.07(5.57)     &  &    2.15(5.94)  &  &     &  &    2.14(5.84)  &  &      &  &      &  &       &  &      \\
				{$^6_{\Lambda}$He}         & 1$^-$    &  &       &  &   $-$3.66   &  &   $-$3.24   &  &   $-$3.35    &  &     4.18$\pm$0.1 &  &   4.42    &  &   4.00     &  &   4.10    \\
				& 2$^-$    &  &        &  &    $-$2.22(0.29)  &  &   $-$2.21(0.30)    &  &   $-$2.94(0.06)   &  &      &  &    2.98  &  &   2.97   &  &    3.69  \\
				& 0$^-$    &  &        &  &    $-$0.95(2.33)  &  &   $-$0.79(2.75)    &  &   $-$0.67(3.71)   &  &      &  &    1.71  &  &   1.55    &  &  1.42    \\
				& 1$^-$    &  &        &  &  $-$0.64(3.79)    &  &   $-$0.72(3.95)    &  &     $-$0.74 (3.62) &  &      &  &   1.40   &  &  1.51     &  &     1.49 \\
				$^6$He         & 0$^+$    &  &     $-$0.975   &  &   $-$0.916   &  &      &  & $-$0.975     &  &      &  &      &  &       &  &      \\
				& 2$^+$    &  &    0.822(0.113)    &  &  0.954(0.176)    &  &        &  & 0.879(0.132)     &  &      &  &      &  &       &  &      \\
				{$^7_{\Lambda}$He}         & 1/2$^+$  &  &        &  &  $-$6.06    &  &  $-$5.70  &  &  $-$6.48    &  &    5.55$\pm$0.10$\pm$0.11   &  &  5.14    &  &    4.78  &  &   5.51   \\
				& 3/2$^+$  &  &       &  &    $-$4.55  &  &   $-$3.95  &  &  $-$4.37    &  &  3.65$\pm$0.20$\pm$0.11    &  &   3.63   &  &    3.03  &  &    3.34  \\
				& 5/2$^+$  &  &        &  &   $-$3.38   &  & $-$3.15      &  & $-$3.84     &  &      &  &   2.46   &  &  2.23     &  &   2.87   \\
				$^7$He          & 3/2$^-$  &  &  $-$0.53(0.15)      &  &  $-$0.33(0.04)    &  &        &  &   $-$0.58(0.048)     &  &      &  &      &  &       &  &      \\
				& 1/2$^-$  &  &   0.37(1.0)      &  &   0.42(1.9)  &  &      &  &  0.42(2.77)    &  &      &  &      &  &       &  &      \\
				{$^8_{\Lambda}$He}         & 1$^-$    &  &      &  &    $-$7.64  &  &    $-$6.76   &  & $-$6.53     &  &  $7.17\pm0.7$    &  &    7.31  &  &   6.43  &  &   5.95   \\
				& 2$^-$   &  &        &  &   $-$6.49   &  &  $-$6.00     &  &   $-$6.33   &  &      &  &   6.16   &  &    5.67  &  &    5.75  \\
				& 0$^-$    &  &        &  &  $-$5.83    &  & $-$5.27      &  & $-$4.27     &  &      &  &  5.50   &  &  4.94     &  &   3.69   \\
				& 1$^-$    &  &        &  &   $-$5.00   &  &    $-$4.61   &  &  $-$4.31    &  &      &  &   4.67  &  &    4.28   &  &  3.73    \\
				$^8$He         & 0$^+$    &  &   $-$3.11     &  &  $-$3.05    &  &      &  &  $-$3.21    &  &      &  &      &  &       &  &      \\
				& 2$^+$    &  &      0.46(0.50)    &  &  0.03(0.24)    &  &      &  &  0.32(0.66)    &  &      &  &      &  &       &  &      \\
				{$^9_{\Lambda}$He}         & 1/2$^+$  &  &        &  &   $-$11.39   &  &   $-$10.54   &  &  $-$10.29    &  &      &  &  8.34    &  &   7.49   &  &   7.09   \\
				& 3/2$^+$  &  &        &  &    $-$8.85  &  &    $-$7.82   &  &  $-$5.69    &  &      &  &   5.80   &  &   4.77  &  &   2.48   \\
				& 5/2$^+$  &  &        &  &   $-$7.14   &  & $-$6.48     &  &   $-$5.53   &  &      &  &  4.09    &  &   3.43  &  &   2.33   \\ \hline \hline
		\end{tabular}}
	\end{table*}

\section{Results and discussion}
	\label{sec:res}

	 The low-energy spectra and resonance widths of the $^A_\Lambda$He hyper-isotopes and their corresponding $^{A{-}1}$ He core nuclei are shown in Fig.~\ref{figure1}. For $^{5-8}$He, one can see that theoretical energies and widths (solid black lines) are in good agreement with experimental data (solid black triangles). The difference between calculated and experimental widths is indeed smaller than 500 keV except the exited state of $^7$He, an error similar to that of Ref. \cite{PhysRevC.96.054316}. In fact, our present model Hamiltonian differ{s} from that of Ref. \cite{PhysRevC.96.054316} only quantitatively, as the parameters of the WS potentials and FHT interaction are fitted to slightly different nuclear states. The He chain is necessary to constrain the $NN$ part of the FHT interaction,  because the available experimental data mostly originate from nuclei rather than hypernuclei. However, it is not possible to determine the $NN$ parameters solely from the He chain, with the coupling constants fitting the remaining hypernuclear states. This paradoxical situation arises because the FHT interaction possesses several sets of parameters that can accurately describe the He chain, but not all of them can be generalized to provide a suitable Hamiltonian for hypernuclei.

		{As previously mentioned, only four experimental binding energies for the considered hypernuclei are available, while the others have not yet been measured. Moreover, these experimental data can bear a large experimental error. For example, only six events for $^8_{\Lambda}$He have been synthesized \cite{davis200550,gal2016strangeness}, with an observed binding energy of 7.17 MeV and a large error bar of 0.7 MeV \cite{davis200550,gal2016strangeness}. This renders the fit analysis and theoretical error assessment even more difficult quantitatively. Nevertheless, we managed to reproduce the observed $\Lambda$ binding energy of $^7_{\Lambda}$He well. As $^5$He is well unbound, the addition of a $\Lambda$ hyperon only makes the ground state of $^6_{\Lambda}$He bound by 540 keV. As a consequence, $^6_{\Lambda}$He is a neutron halo hypernucleus in our calculations. Three excited resonances ($J^\pi$ = 2$^-$, 0$^-$, 1$^-_2$) are predicted above the bound $1_1^-$ ground state, with two exhibiting broad widths of 2.33 and 3.79 MeV (see Fig.~\ref{figure1}). For $^7_{\Lambda}$He, the calculated energy of the ground state is $-$6.06 MeV which corresponds to a $\Lambda$ binding energy of 5.14 MeV. The first 3/2$^+$ state is a bound state and 5/2$^+$ state is a resonance in our model, with binding energies $B_\Lambda$ = 3.63 and 2.46 MeV with respect to the $^6$He + $\Lambda$ threshold, respectively. The JLab E05-115 experiment reports an excited-state peak at $B_\Lambda^{expt}$ = 3.65 $\pm$ 0.20(stat.) $\pm$ 0.11(sys.) MeV but it is not clearly identified \cite{PhysRevC.94.021302}. Our results suggest that this state more likely corresponds to 3/2$^+$. As for $^8_{\Lambda}$He, we obtain two weakly bound states at $B$ = 1.58 MeV (1$^-_1$) and 0.43 MeV (2$^-$), with respect to the $_{\Lambda}^7$He + $n $ threshold. The theoretical prediction of $\Lambda$ binding energy is 7.31 MeV, in good agreement with the experimental value of 7.17 MeV, despite the large experimental error bar. Further experimental data are required to more precisely assess the accuracy of our theoretical predictions. Finally, let us discuss the binding energy of the hypernucleus $^9_{\Lambda}$He. The ground-state energy of the associated core nucleus \( ^8 \text{He} \) is calculated as \( E = -3.05 \) MeV. The addition of the $\Lambda$ hyperon significantly increases the baryonic binding, resulting in a deeply bound state for $^9_{\Lambda}$He with $\Lambda$ binding energy $B_\Lambda$ = 8.34 MeV. {Experimental investigations have reported varying excitation energies and widths for the first excited state of $^8$He, with these discrepancies arising from the application of different scattering reactions. An initial study using proton inelastic scattering identified the first excited state (2$^+$) at 3.57(12) MeV with a width of 0.50(35) MeV {\cite{korsheninnikov1993experimental}}, while a neutron knock-out experiment reported this state's excitation energy as 2.9 $\pm$ 0.2 MeV with a width of 0.3 $\pm$ 0.3 MeV {\cite{markenroth20018he}}. Our theoretical prediction for the excited energy is 3.09 MeV with a width of 0.24 MeV, which is not inconsistent with the average of the measured values.} Due to the attractive nature of the $\Lambda N$ interaction, the 3/2$^+$ state and the 5/2$^+$ state become weakly bound and narrow resonant states, respectively.

			The detailed energy spectra of {the} studied hypernuclei, along with their associated nuclei (i.e.~without the $\Lambda$ particle), are depicted in Table~\ref{table:one}. These are compared with experimental data as well as theoretical predictions from other models. Our calculations demonstrate close agreement between the ground-state hypernuclear binding energies computed by GSM and experimental values, as well as with results obtained by T.~Myo and E.~Hiyama \cite{PhysRevC.107.054302} using the cluster model. However, discrepancies arise in the prediction of low-lying excited states compared to T.~Myo's and E.~Hiyama's results. In particular, for $^6_{\Lambda}$He and $^8_{\Lambda}$He, our analysis reveals a sequence of 1$_1^-$, 2$^-$, 0$^-$, 1$_2^-$ states, which consistent with the $ab$ $initio$ no-core shell model results of R.~Wirth and R.~Roth in Ref. \cite{wirth2018}. In contrast, T.~Myo and E.~Hiyama report a sequence of 1$_1^-$, 2$^-$, 1$_2^-$, 0$^-$ states. When it comes to low-lying excited states, experimental data is available only for the 3/2$^+$ state of $^7_{\Lambda}$He. There is consistency with our computed result of 3.63 MeV, which lies within the experimental error bars. 
			Overall, the results obtained with our model, the cluster approach of Ref. \cite{PhysRevC.107.054302}, and experimental data are qualitatively consistent. The main point of discrepancy concerns the sequence of excited states in $^{6,8}_{\Lambda}$He, where no experimental confirmation is currently available. To improve the predictive power of the model and better constrain its parameters, an increase in the number of experimental data points is essential.

			To assess the influence of $\Lambda N-$$\Sigma N$ couplings on the $\Lambda$ hyperon binding energy in neutron-rich hypernuclei, we also provide the results obtained by excluding the $\Lambda N-$$\Sigma N$ coupling channel (see Table~\ref{table:one}). The $\Lambda$ binding energies of ground state for $^6_\Lambda$He$- ^9_\Lambda$He are reduced by 0.42, 0.36, 0.88 and 0.85 MeV, respectively, highlighting an increasing impact of the coupling as the mass number increases. Although the probabilities of $\Sigma$ hyperon occupation in different partial waves are small,  {on the order of 0.1\%, they are still} sufficient to modify binding energies by about 1 MeV. It is mentioned in Ref. \cite{PhysRevLett.84.3539} that the $\Lambda$$-$$\Sigma$ coupling of the $YN$ potential can be divided into incoherent and coherent parts where the former give a suppression effect while the latter provides an attractive effect. The suppression of incoherent $\Lambda N$$-$$\Sigma N$ coupling effectively solves the overbinding
			problem in $^5_\Lambda $He \cite{PhysRevLett.84.3539}. Furthermore, in the hypernucleus $^3_\Lambda$H, the $\Lambda n-\Sigma^- p$ conversion contributes to the binding energy, primarily due to the attractive Coulomb interaction it creates between the proton and the generated $\Sigma^-$ particle \cite{haidenbauer2020}. In neutron-rich nuclei, the latter conversion also replaces a large fraction of the $T=1$ $NN$ interaction by its $T=0$ counterpart, so that it provides more and more binding energy with increasing neutron number. This can explain the additional binding energy in $^6_\Lambda$He$-$$^9_\Lambda$He.  {Moreover}, the increase is more than twice as large from $^{7}_\Lambda$He to $^{8}_\Lambda$He compared to others, which is probably due to $YN$ pairing. Thus, even if $\Lambda N$$-$$\Sigma N$ coupling generates  {a} small $\Sigma$ occupation probability, it cannot be neglected in practice, as it can augment binding energies by hundreds of keV to a few MeV.

Although the $\Lambda$ particle primarily occupies the $0s_{1/2}$ orbital and has limited influence on the internal structure of well-bound nuclei, its impact is more pronounced in weakly bound systems. In particular, the transformation of unbound $^5$He into a bound $^6_\Lambda$He, and the structural stabilization observed in $^8_\Lambda$He, highlight the role of the $\Lambda$ particle in modifying continuum coupling and suppressing decay widths. These effects contribute to the halo-like character of light hypernuclei near the neutron drip line.

  \begin{table}[htp]
    \centering
    \caption{The calculated rms radii of He isotope and hyper-isotope chains for proton ($R_\mathrm{p}$), neutron ($R_\mathrm{n}$) and hyperon ($R_\mathrm{\Lambda}$).}
    \renewcommand{\arraystretch}{1.5} 
    \setlength{\tabcolsep}{5mm} 
    \begin{tabular}{cccccc}
        \hline \hline 
        & State & $ R_\mathrm{p} $ & $ R_\mathrm{n} $ & $ R_\mathrm{\Lambda} $ \\ \hline
        {$^6_{\Lambda}$He} & 1$^-$  & 1.890 & 2.526 & 2.480 \\
        {$^6$He} & 0$^+$   & & 2.693 & \\
        {$^7_{\Lambda}$He} & 1/2$^+$ & 1.878 & 2.531 & 2.447 \\
        {$^8_{\Lambda}$He} & 1$^-$   & 1.876 & 2.689 & 2.431 \\
        {$^8$He} & 0$^+$   & & 2.860 & \\
        {$^9_{\Lambda}$He} & 1/2$^+$ & 1.857 & 2.660 & 2.377 \\
        \hline \hline
    \end{tabular}\label{table:four}
\end{table}

				\begin{figure}[htp]
				\begin{center}
					\includegraphics[width=7 cm]{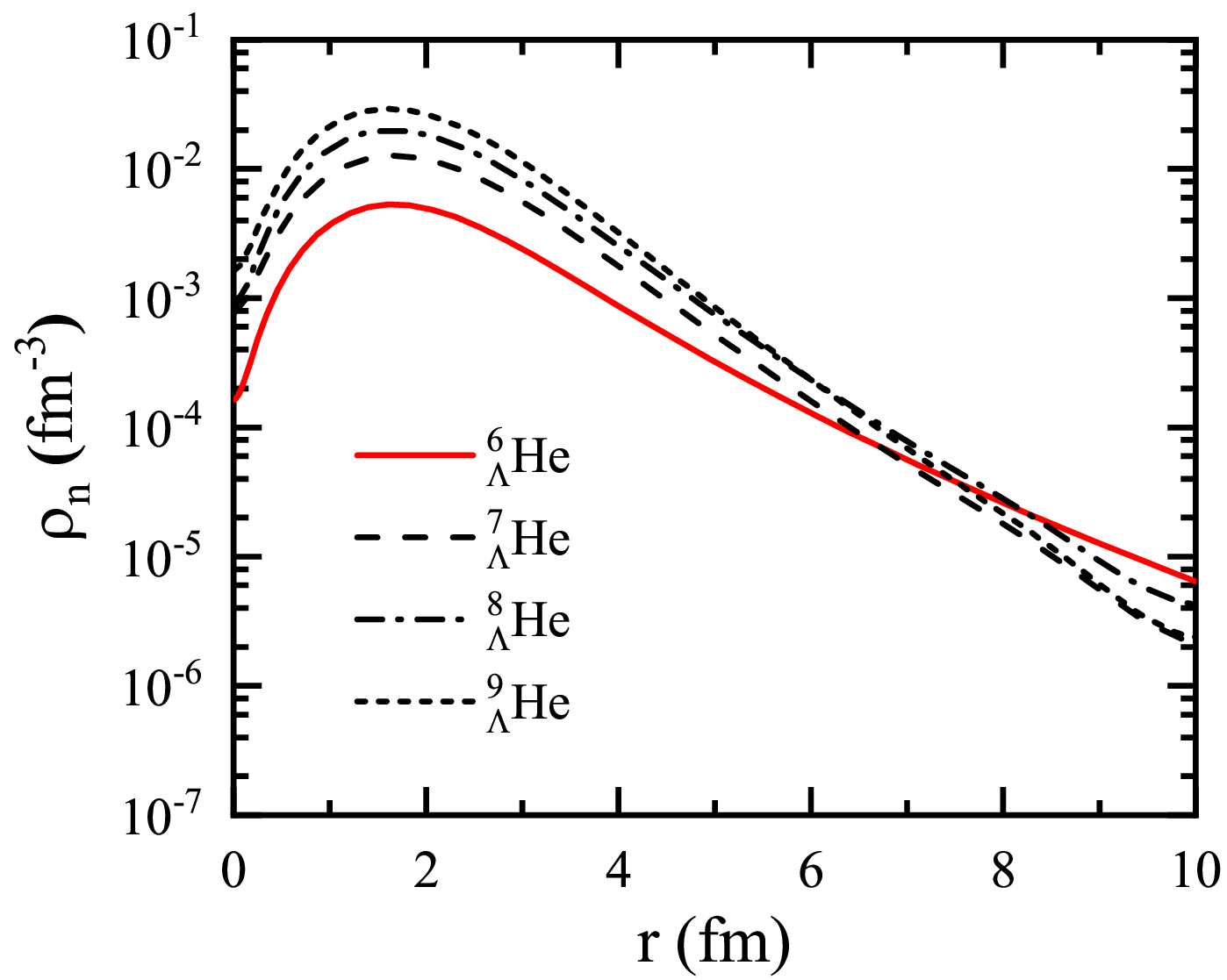}~
					\caption{The neutron density of hypernuclei $_\Lambda^6$He $-$ $ _\Lambda^9$He in their ground states. }\label{figure2}
				\end{center}
			\end{figure}

We also calculated the rms radii of the ground states of the He isotope and hyper-isotope chains (see Table~\ref{table:four}). Since the rms radius is not a meaningful observable for unbound or resonant states, only bound systems are included in the table. The addition of a $\Lambda$ hyperon increases the binding energy and reduces the overall size of the hypernuclei compared to their corresponding nuclei. This glue-like effect is particularly evident in $^6_\Lambda$He and $^8_\Lambda$He, where the core nuclei $^5$He and $^7$He are resonant. In contrast, for $^7_\Lambda$He and $^9_\Lambda$He, the rms radii are only slightly smaller than those of the corresponding $^6$He and $^8$He nuclei, with the contraction induced by the $\Lambda$ particle amounting to approximately 0.2~fm. These findings highlight the role of the $\Lambda$ hyperon in enhancing nuclear stability and inducing modest spatial shrinkage.

			Figure~\ref{figure2} displays the neutron density of hypernuclei in their ground states. Due to the strong dependence of density with respect to neutron separation energy, it directly points out the effect of continuum coupling on wave functions asymptotes. It is particularly visible in $^6_\Lambda$He, which is weakly bound with respect to the $^5_\Lambda$He + $n$ threshold. Indeed, while the density of $^6_\Lambda$He is smaller than that of other hypernuclei at short distance, its decrease is slower than in well bound hypernuclei, so that the density of $^6_\Lambda$He becomes larger than all other neutron densities after 7 fm, whose behavior is sensibly the same. The ground state of $^6_\Lambda$He is a halo state, characterized by a significant spatial extension of the nuclear density compared to well bound systems of similar baryon number.

            		\begin{figure}[htp]
				\begin{center}
					\includegraphics[width=8.5 cm]{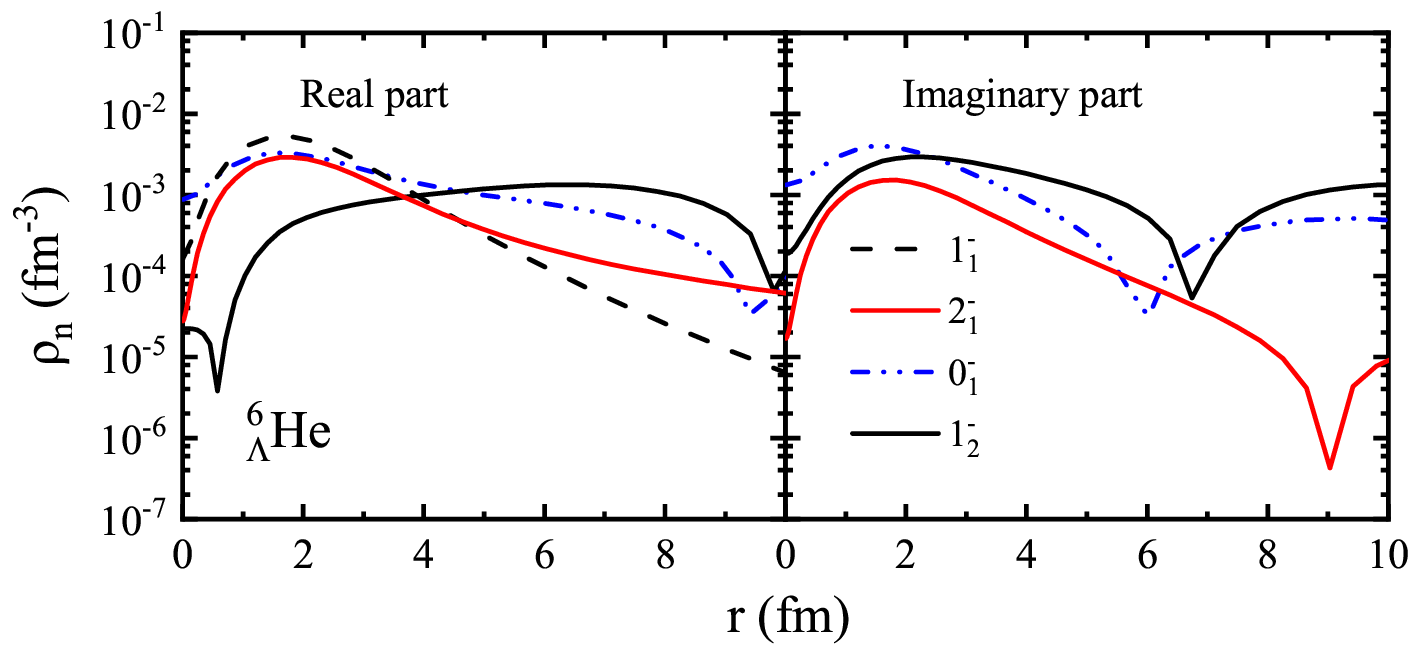}~
					\caption{ The neutron density distribution of the hypernucleus $_\Lambda^6$He for its ground state and low-lying excited states. Imaginary parts not shown are negligible.}\label{figure3}
				\end{center}
			\end{figure}
			
			The real and imaginary parts of the density distributions of both ground {and} excited states of $^6_\Lambda$He are also illustrated in Fig.~\ref{figure3}. Notably, the low-lying excited states 0$^-_1$ and 1$^-_2$ exhibit extended neutron density distributions. Oscillations at large distance also occur in the density of the 0$^-_1$, 1$^-_2$ and 2$^-_1$ hypernuclear states, in both real and imaginary parts for the two former states. This reflects the resonance character of these states. They all bear a sizable width so that their density possesses fairly large real and imaginary parts in absolute value.

%%%%%%%%%%%%%%%%%%%%%%%%%%%%%%%%%%%%%%%%%%%%%%%%%%%%%%%%
%                    begin  summary
%%%%%%%%%%%%%%%%%%%%%%%%%%%%%%%%%%%%%%%%%%%%%%%%%%%%%%%%

\section{Summary}
			\label{sec:sum}

%We have extended the Gamow Shell Model (GSM) to describe neutron-rich hypernuclei, applying it for the first time to the helium hyper-isotopes from $^6_\Lambda$He to $^9_\Lambda$He. The nucleon-nucleon part of the GSM Hamiltonian is adopted from previous GSM studies~\cite{PhysRevC.96.054316,PhysRevC.102.024309}, based on a Woods–Saxon $+$ FHT interaction. The core-$\Lambda$ potential is adjusted to reproduce the experimental binding energy of $^5_\Lambda$He. The hyperon–nucleon interaction is implemented in an effective form with a minimal set of dominant coupling constants tuned to reproduce low-lying states in $^7_\Lambda$He. This approach provides a practical framework for exploring the structure of weakly bound hypernuclei within the GSM framework.

We have extended the Gamow Shell Model (GSM) to investigate weakly bound hypernuclei. The GSM explicitly includes $\Lambda N$–$\Sigma N$ coupling and continuum effects,  which are essential to accurately describe the exotic properties of hypernuclei near the drip line. As an initial application, we apply this framework to study the energy spectra, rms radii, and neutron density distributions of helium hyper-isotopes from $^6_\Lambda$He to $^9_\Lambda$He.

Our GSM calculations reproduce available experimental data for binding energies and low-lying energy spectra of $^6_\Lambda$He–$^8_\Lambda$He. In particular, { we predict a deeply bound state 1/2$^+$, a weakly bound state 3/2$^+$ and a narrow resonance 5/2$^+$ in {$^9_{\Lambda}$He}.} The inclusion of the $\Lambda$ hyperon stabilizes unbound or weakly bound nuclear systems, such as converting the unbound $^5$He and resonant $^7$He nuclei into bound hypernuclei $^6_\Lambda$He and $^8_\Lambda$He, respectively. This effect is seen both in energy spectra and neutron radius reduction.

We also evaluated the impact of the $\Lambda N$–$\Sigma N$ coupling. Though the $\Sigma$ admixture is small, it lowers the binding energies of $^6_\Lambda$He–$^9_\Lambda$He by 0.36–0.88 MeV. This indicates that the inclusion of $\Lambda N-\Sigma N$ channels coupling is crucial for the calculation of hypernuclear energy spectra, and that the common tendency to ignore them is not justified.

In summary, this work provides the first demonstration that the Gamow Shell Model can be successfully extended to hypernuclear systems, offering a reliable description of structure phenomena near the neutron dripline. Although the $YN$ interaction used here is an effective potential, it offers a valuable starting point to explore hypernuclei within this framework. Future developments will incorporate chiral $YN$ interactions and extend to lighter hypernuclei such as $^3_\Lambda$H and $A\sim3$–5 systems.

\section*{Acknowledgements}
			
This work was supported by the National Natural Science Foundation of China
under Grants No.~12175071, No.~12121005, No.~12205340, and No.~12347106, and the Gansu Natural Science Foundation under Grant No.~22JR5RA123;  The numerical calculations in this paper have been done on Hefei advanced computing center.

%%%%%%%%%%%%%%%%%%%%%%%%%%%%%%%%%%%%%%%%%%%%%%%%%%%%%%%%
%                  begin refereee
%%%%%%%%%%%%%%%%%%%%%%%%%%%%%%%%%%%%%%%%%%%%%%%%%%%%%%%%
%%
\bibliographystyle{elsarticle-num-names}
\bibliography{rif}

\end{document}